\pdfoutput=1
\documentclass[prd,onecolumn,notitlepage,eqsecnum,nofootinbib,floatfix]{revtex4-1}
\usepackage{amssymb}
\usepackage{amsmath}
\usepackage{amsfonts} 
\usepackage{bm}
\usepackage{color} 
\usepackage{graphicx}
\DeclareFontFamily{OT1}{rsfs}{} 
\DeclareFontShape{OT1}{rsfs}{m}{n}{<-7> rsfs5 
    <7-10> rsfs7 <10-> rsfs10}{}   
\DeclareMathAlphabet{\scr}{OT1}{rsfs}{m}{n} 
\DeclareSymbolFont{EulerScript}{U}{eus}{m}{n}
\SetSymbolFont{EulerScript}{bold}{U}{eus}{b}{n}
\DeclareSymbolFontAlphabet\scrpt{EulerScript}
 
\newcommand{\B}{{\cal B}} 
\newcommand{\M}{{\scr M}} 
\newcommand{\R}{{\cal R}} 
 
\newcommand{\stf}[1]{{\langle #1 \rangle}} 
\allowdisplaybreaks
\begin{document}
\title{Gravitomagnetic response of an irrotational body to an applied
  tidal field}  
\author{Philippe Landry} 
\affiliation{Department of Physics, University of Guelph, Guelph,
  Ontario, N1G 2W1, Canada} 
\author{Eric Poisson} 
\affiliation{Department of Physics, University of Guelph, Guelph,
  Ontario, N1G 2W1, Canada {\ }} 
\affiliation{Canadian Institute for Theoretical Astrophysics,
  University of Toronto, Toronto, Ontario, M5S 3H8, Canada} 
\date{April 24, 2015} 
\begin{abstract} 
The deformation of a nonrotating body resulting from the application 
of a tidal field is measured by two sets of Love numbers associated
with the gravitoelectric and gravitomagnetic pieces of the tidal
field, respectively. The gravitomagnetic Love numbers were previously
computed for fluid bodies, under the assumption that the fluid is in 
a strict hydrostatic equilibrium that requires the complete absence of
internal motions. A more realistic configuration, however, is an
irrotational state that establishes, in the course of time, internal
motions driven by the gravitomagnetic interaction. We recompute the
gravitomagnetic Love numbers for this irrotational state, and show
that they are dramatically different from those associated with the
strict hydrostatic equilibrium: While the Love numbers are positive in
the case of strict hydrostatic equilibrium, they are negative in the 
irrotational state. Our computations are carried out in the context of 
perturbation theory in full general relativity, and in a
post-Newtonian approximation that reproduces the behavior of the Love 
numbers when the body's compactness is small.   
\end{abstract} 
\maketitle

\section{Introduction and summary} 
\label{sec:intro} 

A body subjected to an applied tidal field suffers a deformation that
depends on the details of its internal structure. When the body is
nonrotating, these details are encapsulated in a set of gravitational 
Love numbers $K^{\rm el}_\ell$ and $K^{\rm mag}_\ell$, and a
measurement of the tidal properties of a body can reveal, through the
Love numbers, important information regarding this internal
structure. This observation has motivated the development of a
relativistic theory of tidal deformation and dynamics, in the context
of the measurement of tidal effects in gravitational waves emitted by
neutron-star binaries \cite{flanagan-hinderer:08, hinderer:08,
  hinderer-etal:10, baiotti-etal:10, baiotti-etal:11,
  vines-flanagan-hinderer:11, pannarale-etal:11, lackey-etal:12,
  damour-nagar-villain:12, read-etal:13, vines-flanagan:13,
  lackey-etal:14, favata:14, yagi-yunes:14} and during the capture of
solar-mass compact bodies by supermassive black holes \cite{hughes:01,
  price-whelan:01, martel:04, yunes-etal:10, yunes-etal:11,
  chatziioannou-poisson-yunes:13}. Tidal invariants have been
incorporated in point-particle actions to account for the tidal
response of an extended body \cite{bini-damour-faye:12, 
  chakrabarti-delsate-steinhoff:13a,
  chakrabarti-delsate-steinhoff:13b, dolan-etal:14, bini-damour:14}. 

While the response of a self-gravitating body to an applied
gravitoelectric tidal field is familiar from Newtonian theory (see,
for example, Sec.~2.5 of Ref.~\cite{poisson-will:14} for a thorough
treatment), its response to a gravitomagnetic tidal field is a
relativistic effect that has no analogue in Newtonian gravity. This
effect was first explored by Favata \cite{favata:06} in the context of
post-Newtonian theory, and subsequently by Damour and 
Nagar \cite{damour-nagar:09} and Binnington and Poisson
\cite{binnington-poisson:09} in full general relativity. We examine 
it further in this work, and inspired by Favata, we lift an important
restriction on the types of fluid configurations that were allowed in
the earlier, fully relativistic work.  

The gravitomagnetic Love numbers $K^{\rm mag}_\ell$ of a fluid body
were computed by Damour and Nagar and Binnington and Poisson under the
assumption that the tidal interaction is sufficiently slow that it
never takes the body out of hydrostatic equilibrium. This is a
good approximation for many circumstances; for example, it is expected
to hold for most of the orbital evolution of a compact binary system,
up to the point where merger is about to take place. But the
hydrostatic equilibrium considered in the earlier work is a strict one
that forbids the existence of fluid motions within the body; the
compact body is assumed to be strictly static, except for the
parametric time dependence communicated by the slowly changing state  
of the tidal environment.      

Our main purpose in this paper is to point out that the strict
hydrostatic equilibrium is too severe a restriction on the body's
internal physics. We follow instead Shapiro \cite{shapiro:96} and
Favata \cite{favata:06}, and take the fluid to be in an irrotational
state that permits internal motions driven by the gravitomagnetic
interaction with the tidal environment. We recalculate the Love
numbers $K^{\rm mag}_\ell$ for this configuration, and show that they
are dramatically different from those associated with the strict
hydrostatic equilibrium: While the Love numbers are positive in the
case of strict hydrostatic equilibrium, they are negative in the
irrotational state. 

In our work the tidal field is still taken to vary slowly, and the
fluid is still taken to be in an approximate hydrostatic equilibrium,
in the sense that the fluid's physical variables, such as density,
pressure, and velocity field, carry only a parametric dependence upon
time that reflects the slow evolution of the tidal environment. But
internal motions are now allowed. As Shapiro and Favata have shown,
these internal motions are a consequence of the conservation of
relativistic circulation within the fluid, and are established
whenever the tidal field exhibits a time dependence, however slow it
may happen to be. On the other hand, the strict hydrostatic
equilibrium adopted in the earlier work requires the tidal environment
to be strictly stationary; it is a far less realistic description of
the fluid. Our considerations in this paper are limited to the
gravitomagnetic interaction; as we shall show, the switch from strict
hydrostatic equilibrium to the irrotational state has no impact on the
body's gravitoelectric response.  

We begin our developments in Sec.~\ref{sec:unperturbed} with a
description of the unperturbed state of an isolated, self-gravitating
body consisting of a perfect fluid; we take the unperturbed
configuration to be static and spherically symmetric. In
Sec.~\ref{sec:perturbed} we introduce a perturbation and examine the
relativistic Euler equation that governs the perturbed state of the
fluid. We continue the discussion in Sec.~\ref{sec:irrotational} by
working out the consequences of the relativistic circulation theorem
for our perturbed configuration; we show that the irrotational state
comes with internal motions that are forbidden in the strict
hydrostatic equilibrium.  

In Sec.~\ref{sec:Love} we specialize the perturbation to a
gravitomagnetic tidal field, and we calculate the body's response to
this field when the fluid configuration is in the irrotational
state. The tidal environment is generic and characterized by an 
$\ell$-pole gravitomagnetic moment $\B_{k_1 k_2 \cdots k_\ell}(t)$
that is assumed to vary slowly with time. This Cartesian tensor is
symmetric and tracefree (STF), and in a quasi-Lorentzian frame 
$(t, x^j)$ it appears in the time-space components of the metric
tensor,   
\begin{equation} 
g_{tj} = \frac{2}{3(\ell-1)} \epsilon_{jpq} 
\B^q_{\ k_2 k_3 \cdots k_\ell} x^p x^{k_2} x^{k_3} \cdots x^{k_\ell} 
\biggl[ 1 + \cdots - 2 \frac{\ell+1}{\ell} K^{\rm mag}_\ell   
\biggl( \frac{2M}{r} \biggr)^{2\ell+1} (1 + \cdots) \biggr]. 
\label{g0j} 
\end{equation} 
Here, $\epsilon_{jpq}$ is the completely antisymmetric permutation
symbol, $K^{\rm mag}_\ell$ is the gravitomagnetic Love number of
degree $\ell \geq 2$, $M$ is the body's gravitational mass, 
$r^2 := \delta_{jk} x^j x^k$, and dots indicate relativistic
corrections of order $2M/r$ and higher; we work in relativistic units
with $c = G = 1$. This expression for $g_{tj}$ applies to a domain 
$R < r < r_{\rm out} \ll b$ bounded internally by the body's radius
$R$ and externally by an outer radius $r_{\rm out}$ required to be
much smaller than $b$, the distance to the external matter responsible
for the tidal field. The first term in $g_{tj}$, which grows as
$r^\ell$, represents the external tidal field, and the second term,
which decays as $r^{-\ell-1}$, represents the body's response to the
applied tidal field, quantified by $K^{\rm mag}_\ell$. The tidal
moments $\B_{k_1 k_2 \cdots k_\ell} (t)$ can be thought of as a
collection of functions of time that cannot be determined by the
Einstein field equations restricted to the domain 
$R < r < r_{\rm out}$; or they can be viewed as components of the
Riemann tensor differentiated $\ell-2$ times and evaluated in the
regime $r \gg M$,  
\begin{equation} 
\B_{k_1 k_2 \cdots k_\ell} = \frac{3}{2(\ell+1)(\ell-2)!!} 
\bigl( \epsilon_{k_1 p q} R^{pq}_{\ \ k_2 0 ; k_3 \cdots k_\ell}
\bigr)^{\rm STF}, 
\end{equation} 
in which the STF label instructs us to symmetrize the $k_1 k_2 \cdots   
k_\ell$ indices and remove all traces. This notation, and our
expression for the external piece of $g_{0j}$, is imported from
Zhang's pioneering work \cite{zhang:86}. For a generic tidal 
environment the dominant moment is $\B_{jk}$, and the most
relevant Love number is $K^{\rm mag}_2$, but a formulation of the
body's tidal response can be provided for any multipole order.  

\begin{figure} 
\includegraphics[width=1.0\linewidth]{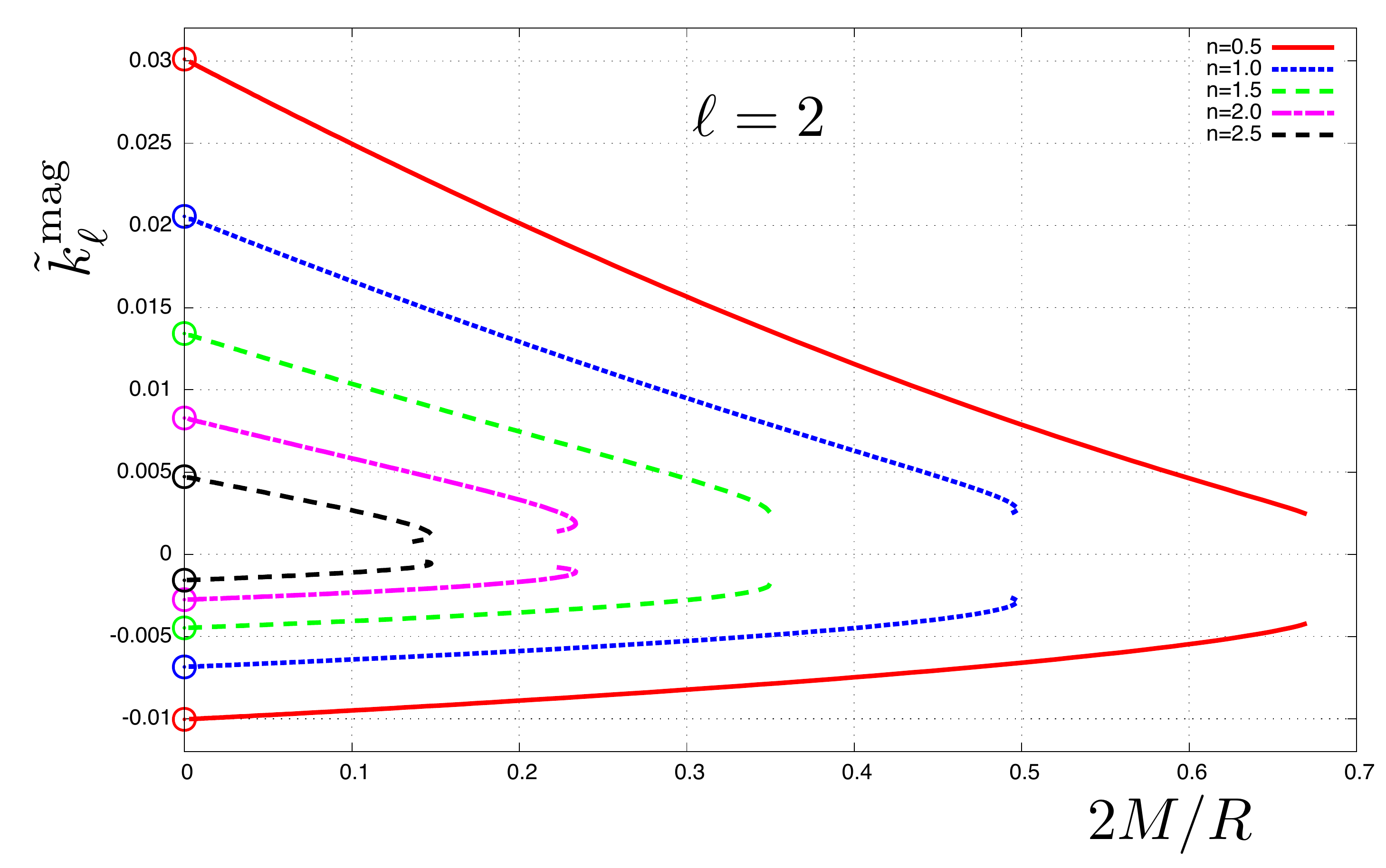}
\caption{Rescaled gravitomagnetic Love number 
$\tilde{k}^{\rm mag}_\ell$ for $\ell = 2$. The Love numbers are
computed for selected polytropes with polytropic index $n$, and
plotted as a function of $2M/R$, up to (and slightly beyond when 
$n > 1$) the maximum value for the given equation of state. The
positive values correspond to the strict hydrostatic equilibrium. The
negative values correspond to the irrotational state. The circled data
points at $2M/R = 0$ are the post-Newtonian values displayed in
Eq.~(\ref{kPN}).}    
\label{fig:fig1} 
\end{figure} 

The gravitomagnetic Love numbers $K^{\rm mag}_\ell$ depend on the
details of the body's internal structure, as determined by its
equation of state, which we take to be of the zero-temperature form 
$p = p(\rho)$, $\epsilon = \epsilon(\rho)$, where $\rho$ is the
rest-mass density, $p$ the pressure, and $\epsilon$ the density of
internal energy. For concreteness we adopt a simple polytropic model   
$p = K \rho^{1+1/n}$, $\epsilon = n p$, where $K$ and $n$ are
constants. A sample of our computations is displayed in
Fig.~\ref{fig:fig1}, which plots 
\begin{equation} 
\tilde{k}^{\rm mag}_\ell :=  \biggl( \frac{2M}{R} \biggr)^{2\ell}   
K_\ell^{\rm mag} 
\end{equation} 
as a function of $2M/R$ for $\ell = 2$ and selected values of the
polytropic index $n$. We observe that the tidal response of an
irrotational fluid is dramatically different from the response of a
fluid in strict hydrostatic equilibrium: the gravitomagnetic Love
numbers of an irrotational body are negative, while they are positive
for a body in strict hydrostatic equilibrium. Both sets of Love
numbers, however, share  the properties that they decrease (in  
absolute value) with increasing $2M/R$ and with increasing $n$;
because increasing $n$ decreases $p$ for a given $\rho$, and therefore
produces a body that is more centrally dense, both properties are
associated with the fact that a more compact body develops smaller
multipole moments.  

In Sec.~\ref{sec:PN} we exploit the methods of post-Newtonian theory
to calculate $\tilde{k}^{\rm mag}_\ell$ in the limit $2M/R \to 0$, for
any equation of state. We obtain the simple expression 
\begin{equation} 
\tilde{k}^{\rm mag}_\ell \to 
-\frac{2 \bigl[(2\lambda-1)\ell - 1 \bigr]}{(\ell+1)(2\ell+1)} 
\frac{\pi}{M R^{2\ell}} \int_0^R \rho r^{2\ell+2}\, dr,  
\label{kPN} 
\end{equation} 
which can be evaluated explicitly once the equation of state is
specified. The parameter $\lambda$ tracks the internal motions
associated with the fluid's irrotational state; setting $\lambda = 1$  
places the body in the irrotational state and gives rise to a negative
$\tilde{k}^{\rm mag}_\ell$ for any $\ell$ and any equation of state, 
while setting $\lambda = 0$ places the body in a strict hydrostatic
equilibrium and gives rise to a positive gravitomagnetic Love number. 
The post-Newtonian values appear as circled data points in
Fig.~\ref{fig:fig1}, and we see that they accurately reproduce the
$2M/R \to 0$ limit of the relativistic curves.   

We conclude this introduction with a restatement of our main
message. A nonrotating body in a tidal interaction with remote 
matter is expected to be in an irrotational state for which internal
motions get established over time. These internal motions have a
dramatic influence on the body's gravitomagnetic Love numbers:   
while they are positive in the strict hydrostatic equilibria examined
in previous works, they are negative in the irrotational state. This
effect is captured by the post-Newtonian expression of
Eq.~(\ref{kPN}), which is valid in the limit $2M/R \to 0$ for any
equation of state.  
  
\section{Unperturbed configuration} 
\label{sec:unperturbed} 

The unperturbed body is taken to be static and spherically
symmetric, and to consist of a perfect fluid with rest-mass density
$\rho$, pressure $p$, density of internal energy $\epsilon$, and
density of total energy $\mu = \rho + \epsilon$. The background
spacetime has a metric given by  
\begin{equation} 
ds^2 = -e^{2\psi}\, dt^2 + f^{-1}\, dr^2 + r^2\, d\Omega^2 
\label{g_back} 
\end{equation} 
with $f := 1-2m/r$, in which $\psi$ and $m$ depend on the 
radial coordinate $r$, and $d\Omega^2 := d\theta^2 
+ \sin^2\theta\, d\phi^2$. The metric is a solution to the Einstein
field equations with energy-momentum tensor 
\begin{equation} 
T^{\alpha\beta} = (\mu+p) u^\alpha u^\beta + p g^{\alpha\beta}, 
\label{em_tensor} 
\end{equation} 
in which $u^\alpha$ is the fluid's velocity field, with only
nonvanishing component $u^t = e^{-\psi}$. The metric functions are
determined by   
\begin{equation} 
m' = 4\pi r^2 \mu, \qquad 
\psi' =\frac{m + 4\pi r^3 p}{r^2 f}, 
\label{EFE_back} 
\end{equation} 
in which a prime indicates differentiation with respect to $r$. In the
vacuum exterior the field equations produce the Schwarzschild solution
$m = M = \mbox{constant}$ and $e^{2\psi} = 1-2M/r$. The body's surface
is situated at $r=R$, as determined by the condition $p(R) = 0$. 

The conservation equation $\nabla_\beta T^{\alpha\beta} = 0$ gives
rise to the relativistic Euler equation, 
\begin{equation} 
(\mu + p) a_\alpha + \bigl( \delta_\alpha^{\ \beta} 
+ u_\alpha u^\beta \bigr) \nabla_\beta p = 0,  
\label{Euler} 
\end{equation} 
where $a^\alpha := u^\beta \nabla_\beta u^\alpha$ is the covariant
acceleration, and the first law of thermodynamics, 
\begin{equation} 
u^\alpha \nabla_\alpha \mu + (\mu + p) \nabla_\alpha u^\alpha = 0. 
\label{first_law} 
\end{equation} 
For the unperturbed configuration the only nonvanishing component of
the acceleration vector is $a_r = \psi'$, and Eq.~(\ref{Euler})
reduces to  
\begin{equation} 
p' = -(\mu+p) \psi' = -\frac{(\mu+p)(m + 4\pi r^3 p)}{r^2 f},  
\label{TOV}
\end{equation} 
the Tolman-Oppenheimer-Volkov equation.  

\section{Perturbed configuration} 
\label{sec:perturbed} 

We next allow the fluid configuration to be perturbed by a
time-dependent, external tidal field. The metric becomes
$g_{\alpha\beta} + p_{\alpha\beta}$, and the fluid quantities are
shifted to $\mu + \delta \mu$, $p + \delta p$, 
$u^\alpha + \delta u^\alpha$, $u_\alpha + \delta u_\alpha$,  
and $a_\alpha + \delta a_\alpha$. We work consistently to first order 
in the perturbation, and note that $\delta u_\alpha 
= g_{\alpha\beta} \delta u^\beta + p_{\alpha\beta} u^\beta$.  

Normalization of the perturbed velocity vector in the perturbed metric
implies that $u_\alpha \delta u^\alpha = -\frac{1}{2} p_{\alpha\beta}
u^\alpha u^\beta$, or $\delta u^t = \frac{1}{2} e^{-3\psi}
p_{tt}$. The remaining components are denoted $\delta u^r := v^r$ and
$\delta u^A := v^A$, with $\theta^A = (\theta,\phi)$. We let 
$v_\alpha := g_{\alpha\beta} v^\beta$ and find that the components of 
$\delta u_\alpha$ are given by  
\begin{equation} 
\delta u_t = \frac{1}{2} e^{-\psi} p_{tt}, \qquad 
\delta u_r = v_r + e^{-\psi} p_{tr}, \qquad 
\delta u_A = v_A + e^{-\psi} p_{tA}. 
\label{delta-u} 
\end{equation} 
A straightforward computation further reveals that 
\begin{subequations} 
\label{delta-a} 
\begin{align}  
\delta a_t &= -e^\psi \psi' v^r, \\
\delta a_r &= e^{-\psi}\partial_t \bigl( v_r + e^{-\psi} p_{tr} \bigr) 
- \frac{1}{2} \partial_r \bigl( e^{-2\psi} p_{tt} \bigr), \\ 
\delta a_A &=e^{-\psi} \partial_t \bigl( v_A + e^{-\psi} p_{tA} \bigr) 
- \frac{1}{2} \partial_A \bigl( e^{-2\psi} p_{tt} \bigr). 
\end{align} 
\end{subequations} 

The perturbed configuration is governed by Euler's equation, which
becomes  
\begin{equation} 
(\mu+p) \delta a_\alpha + (\delta\mu + \delta p) a_\alpha 
+ \bigl( \delta_\alpha^{\ \beta} + u_\alpha u^\beta \bigr) 
  \nabla_\beta \delta p 
+ \bigl( u_\alpha \delta u^\beta + u^\beta \delta u_\alpha \bigr) 
  \nabla_\beta p = 0
\end{equation} 
after the perturbation. With the information provided above, we find
that the radial component reads  
\begin{equation} 
e^{-\psi} (\mu+p) \partial_t \bigl( v_r + e^{-\psi} p_{tr} \bigr) 
- \frac{1}{2} (\mu+p) \partial_r \bigl( e^{-2\psi} p_{tt} \bigr) 
+ (\delta \mu + \delta p) \psi' + \partial_r \delta p = 0, 
\label{Euler_radial} 
\end{equation} 
while the angular components take the form of 
\begin{equation} 
e^{-\psi} (\mu+p) \partial_t \bigl( v_A + e^{-\psi} p_{tA} \bigr)  
+ \partial_A \biggl[ -\frac{1}{2} (\mu+p) e^{-2\psi} p_{tt} 
+ \delta p \biggr] = 0. 
\label{Euler_angular} 
\end{equation} 
The time component of Euler's equation returns a trivial $0=0$.   

\section{Irrotational configuration} 
\label{sec:irrotational} 

The circulation of a relativistic fluid around a closed curve $c$ is
defined by  
\begin{equation} 
C(c) := \oint_c h u_\alpha\, dx^\alpha, 
\label{C-def} 
\end{equation} 
where $h$ is the specific enthalpy defined by 
$d\ln h := (\mu+p)^{-1}\, dp$, and $dx^\alpha$ is the coordinate
increment along $c$. It is known that $C$ is the same for {\it any}
circuit $c$ that surrounds a given fluid world tube. (The world tube
can be thought of as a bundle of streamlines, defined as the world
lines of fluid elements.) Thus, if $c_1$ surrounds the world tube at a 
time $t=t_1$, and if $c_2$ surrounds the same world tube at a time
$t=t_2$, then $C(c_2) = C(c_1)$ and the circulation is conserved. A
proof of the circulation theorem can be found in Synge's 1937 review
of relativistic hydrodynamics \cite{synge:37}.  

Because $u_\alpha\, dx^\alpha = 0$ for the unperturbed configuration,
we have that the unperturbed fluid is irrotational: $C(c) = 0$ for any
purely spatial circuit $c$. The circulation of the perturbed
configuration is then  
\begin{equation} 
C(c) = \oint_c \bigl( \delta h u_\alpha
+ h \delta u_\alpha \bigr) dx^\alpha   
= \oint_c h \delta u_\alpha\, dx^\alpha. 
\end{equation} 
We assume that the fluid begins in an unperturbed state, so that the
perturbation vanishes at some initial time $t = t_0$. This implies
that $C(c_0) = 0$ for any circuit $c_0$ tangent to the hypersurface 
$t = t_0$. Conservation of circulation then guarantees that the
perturbed fluid is irrotational at all times: $C(c_t) = 0$ for any
circuit $c_t$ tangent to any hypersurface $t = \mbox{constant} 
> t_0$. And because the integral of $h \delta u_\alpha\, dx^\alpha$   
must vanish for any circuit $c_t$, we conclude that an irrotational
fluid configuration must satisfy  
\begin{equation} 
\delta u_r = 0 = \delta u_A 
\end{equation} 
at all times. Importing Eq.~(\ref{delta-u}), we have that  
\begin{equation} 
v_r + e^{-\psi} p_{tr} = 0, \qquad 
v_A + e^{-\psi} p_{tA} = 0 
\label{irrot} 
\end{equation} 
for an irrotational configuration.  

We next insert the second of Eqs.~(\ref{irrot}) within 
Eq.~(\ref{Euler_angular}). The equation integrates to  
\begin{equation} 
\delta p = \frac{1}{2} (\mu+p) e^{-2\psi} p_{tt}, 
\end{equation} 
and this can be substituted back into Eq.~(\ref{Euler_radial}), along
with the first of Eqs.~(\ref{irrot}). This yields 
\begin{equation} 
\psi' \delta \mu =  -\frac{1}{2} e^{-2\psi} \mu' p_{tt}. 
\end{equation} 
With the help of Eq.~(\ref{TOV}) these expressions become 
\begin{equation} 
\delta \mu = - r\mu' F, \qquad 
\delta p = -r p' F, \qquad 
F := \frac{p_{tt}}{2 r e^{2\psi} \psi'}.  
\label{gen_hydro} 
\end{equation} 
These equations imply that a spherical surface of constant $\mu$ or
$p$ at radius $r$ in the unperturbed configuration is deformed by the
perturbation to a nonspherical surface at $r(1+F)$. Equations 
(\ref{gen_hydro}) are a consequence of the perturbed Euler equation
and the conservation of circulation for an irrotational fluid. They
allow the perturbation $p_{\alpha\beta}$ to be time-dependent, but
they are identical in form to the equations of hydrostatic equilibrium
derived, for example, by Landry and Poisson \cite{landry-poisson:14}. 

We are interested in the deformation of a fluid body placed in a tidal
environment, and we shall henceforth assume that the tidal field
varies slowly with time. In this context, the state of the fluid is at 
all times an approximate hydrostatic equilibrium described by
Eqs.~(\ref{gen_hydro}), in which all quantities vary slowly with
time. The fluid, however, is not taken to be in a strict hydrostatic  
equilibrium, which would imply the complete absence of internal 
motions. Instead, the fluid is assumed to be in an irrotational state,
which implies the existence of a velocity field described by
Eq.~(\ref{irrot}). The internal motions are established even when the
time dependence of the tidal field is arbitrarily slow. By contrast,
the strict hydrostatic equilibrium previously studied by Damour and
Nagar \cite{damour-nagar:09} and Binnington and Poisson
\cite{binnington-poisson:09} can only be established when the tidal
field is strictly time-independent. As such, it represents a much less
realistic configuration for the perturbed fluid.      

\section{Gravitomagnetic Love numbers of an irrotational compact body} 
\label{sec:Love} 

In this section we determine the impact of the internal motions
described by Eq.~(\ref{irrot}) on the body's response to an applied
tidal field, as measured by the gravitomagnetic Love numbers 
$K^{\rm mag}_\ell$; we shall show that they have no impact on the
body's gravitoelectric response. We assume that the tidal environment
varies slowly with time, and neglect all time derivatives in the
equations that determine the metric perturbation
$p_{\alpha\beta}$. But while the slow time dependence is assumed not
to have an impact on the field equations, it is crucial in the
establishment of the internal motions, as was explained in the
preceding section.   

To proceed we introduce a decomposition of $p_{\alpha\beta}$ into
tensorial spherical harmonics. Denoting $x^a := (t,r)$ and 
$\theta^A = (\theta,\phi)$, we have 
\begin{subequations} 
\label{sphharm_decomp} 
\begin{align} 
p_{ab} &= \sum_{\ell m} h^{\ell m}_{ab} Y^{\ell m}, \\ 
p_{aB} &= \sum_{\ell m} j_a^{\ell m} Y_A^{\ell m}
+ \sum_{\ell m} h_a^{\ell m} X_A^{\ell m}, \\ 
p_{AB} &= r^2 \sum_{\ell m} \bigl( K^{\ell m} \Omega_{AB} Y^{\ell m} 
+ G^{\ell m} Y_{AB}^{\ell m} \bigr) 
+ \sum_{\ell m} h_2^{\ell m} X_{AB}^{\ell m}, 
\end{align} 
\end{subequations} 
in which $h^{\ell m}_{ab}$, $j_a^{\ell m}$, $h_a^{\ell m}$, 
$K^{\ell m}$, $G^{\ell m}$, and $h^{\ell m}_2$ depend on $x^a$ only, and  
\begin{subequations} 
\label{tensor_harmonics} 
\begin{align} 
Y_A^{\ell m} &:= D_A Y^{\ell m}, \\ 
Y_{AB}^{\ell m} &:= \biggl[ D_A D_B + \frac{1}{2} \ell(\ell+1)
\Omega_{AB} \biggr] Y^{\ell m}, \\  
X_A^{\ell m} &:= -\varepsilon_A^{\ B} D_B Y^{\ell m}, \\ 
X_{AB}^{\ell m} &:= \frac{1}{2} \bigl( D_A X_B^{\ell m}
+ D_B X_A^{\ell m} \bigr) 
\end{align} 
\end{subequations} 
are the vectorial and tensorial harmonics constructed from the
standard spherical harmonics $Y^{\ell m}(\theta^A)$; 
$\Omega_{AB} := \mbox{diag}(1,\sin^2\theta)$ is the metric on a unit
two-sphere, $D_A$ is the covariant derivative operator compatible with
this metric, $\varepsilon_{AB}$ is the Levi-Civita tensor on the unit
two-sphere, with components $\varepsilon_{\theta\phi} 
= -\varepsilon_{\phi\theta} = \sin\theta$, and upper-case Latin
indices are raised with $\Omega^{AB}$, the matrix inverse to
$\Omega_{AB}$. The terms involving $Y^{\ell m}$, $Y_A^{\ell m}$, and 
$Y_{AB}^{\ell m}$ in Eq.~(\ref{sphharm_decomp}) constitute the
even-parity sector of the perturbation, and the terms involving
$X^{\ell m}_A$ and $X_{AB}^{\ell m}$ constitute the odd-parity
sector. 

The body's response to the applied tidal field is measured by the Love 
numbers $K^{\rm el}_\ell$ and $K^{\rm mag}_\ell$ introduced by
Binnington and Poisson \cite{binnington-poisson:09}. Because these are 
gauge-invariant in the usual sense of perturbation theory, we may
calculate them in any gauge, and for this purpose it is convenient to
adopt the Regge-Wheeler gauge, for which $j_a^{\ell m}$, $G^{\ell m}$,
and $h_2^{\ell m}$ are all set equal to zero. The field equations
further imply that $h_{tr}^{\ell m} = h_r^{\ell m} = 0$, and we find
that Eqs.~(\ref{irrot}) become    
\begin{equation} 
v_r = 0, \qquad 
v_A = -\lambda e^{-\psi} p_{tA} 
= -\lambda e^{-\psi} \sum_{\ell m} h_t^{\ell m} X_A^{\ell m} 
\label{irrot2} 
\end{equation} 
in Regge-Wheeler gauge. This implies that the internal motions
associated with the irrotational state affect only the odd-parity
sector of the perturbation. This, in turn, implies that the body's
gravitoelectric response to the tidal field is unaffected by the
internal motions, but that there is an impact on the gravitomagnetic
response. In other words, the internal motions do influence the 
gravitomagnetic Love numbers $K^{\rm mag}_\ell$, but they 
leave the gravitoelectric Love numbers $K^{\rm el}_\ell$ unchanged 
with respect to a strict hydrostatic equilibrium. We have inserted a
factor $\lambda \equiv 1$ in Eq.~(\ref{irrot2}) to track the impact of
the internal motions on the computation of $K^{\rm mag}_\ell$; setting
$\lambda = 0$ would place the fluid in a strict hydrostatic
equilibrium.   

The computation of the gravitomagnetic Love numbers proceeds as 
detailed in Ref.~\cite{binnington-poisson:09} and streamlined in
Ref.~\cite{landry-poisson:14}. The only change concerns the additional  
contribution to $\delta T_{\alpha\beta}$, the perturbation of the
fluid's energy-momentum tensor, created by the internal motions
described by Eq.~(\ref{irrot2}). Focusing on the odd-parity sector,
writing  
\begin{equation} 
\delta T_{\alpha\beta} = (\mu+p) ( u_\alpha \delta u_\beta 
+ u_\beta \delta u_\alpha) + p p_{\alpha\beta}, 
\end{equation}   
and making use of Eqs.~(\ref{delta-u}) and (\ref{irrot2}), we find
that the only relevant component is 
\begin{equation} 
\delta T_{tA} = \bigl[ (\lambda-1) \mu + \lambda p \bigr] 
\sum_{\ell m} h_t^{\ell m} X_A^{\ell m}. 
\end{equation} 
With this, the Einstein field equations linearized about the
unperturbed solution of Sec.~\ref{sec:unperturbed} imply that the
perturbation fields $h_t^{\ell m}(r)$ are determined by   
\begin{equation} 
r^2 \frac{d^2 h_t^{\ell m}}{dr^2} - P r \frac{d h_t^{\ell m}}{dr} 
- Q h_t^{\ell m} = 0, 
\label{deq1} 
\end{equation} 
where 
\begin{subequations} 
\begin{align} 
P &:= 4\pi r^2 f^{-1} (\mu+p), \\ 
Q &:= f^{-1} \bigl[ \ell(\ell+1) - 4m/r 
- (2\lambda-1) 8\pi r^2 (\mu+p) \bigr]. 
\end{align} 
\end{subequations} 
We observe that the switch from strict hydrostatic equilibrium to an
irrotational state changes the sign of the $8\pi r^2 (\mu+p)$ term in
$Q$. 

\begin{figure} 
\includegraphics[width=1.0\linewidth]{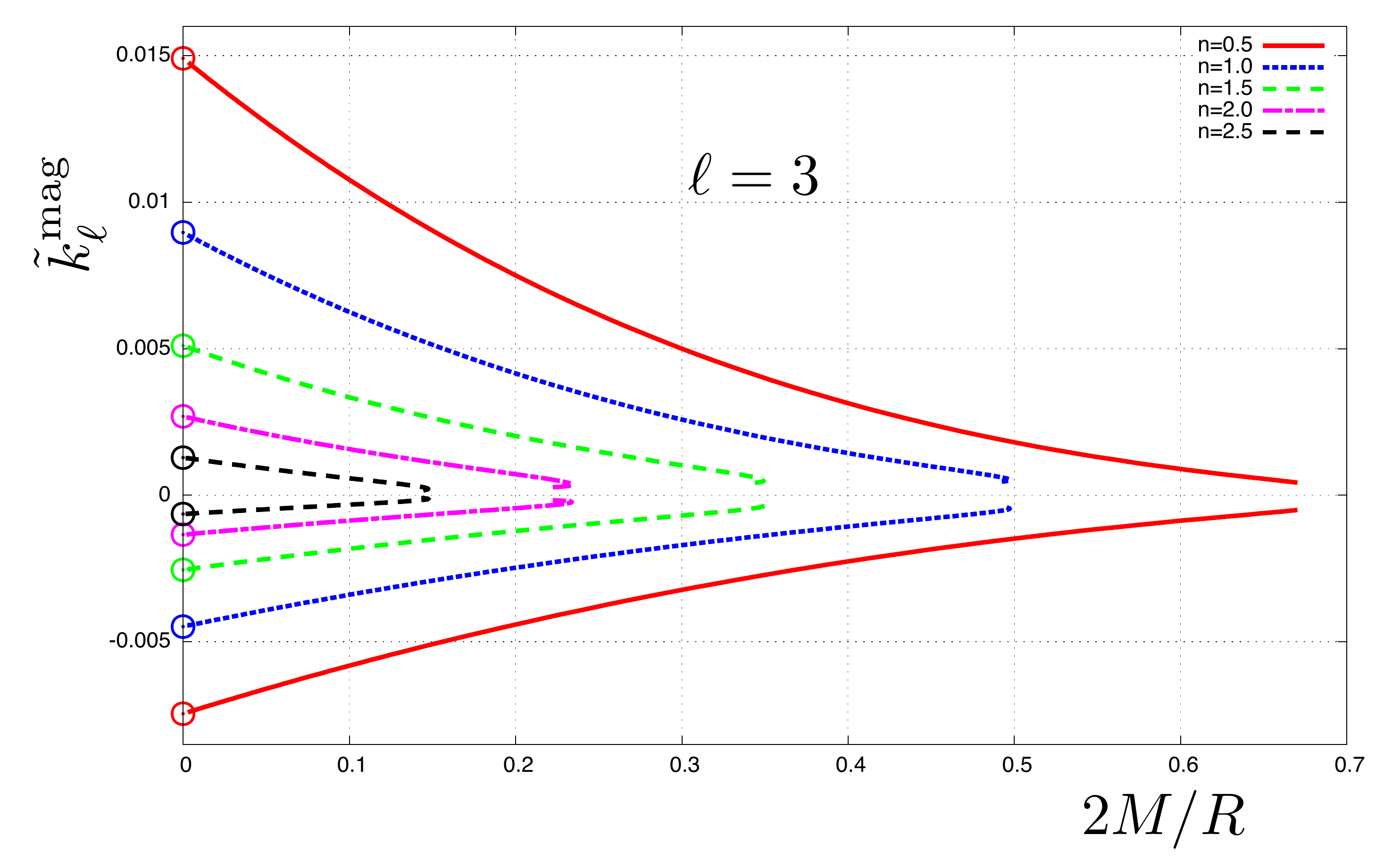}
\caption{Rescaled gravitomagnetic Love number 
$\tilde{k}^{\rm mag}_\ell$ for $\ell = 3$. The Love numbers are 
computed for selected polytropes with polytropic index $n$, and
plotted as a function of $2M/R$, up to (and slightly beyond when 
$n > 1$) the maximum value for the given equation of state. The
positive values correspond to $\lambda = 0$, the strict hydrostatic
equilibrium. The negative values correspond to $\lambda = 1$, the
irrotational state. The circled data points at $2M/R = 0$ are the
post-Newtonian values obtained in Sec.~\ref{sec:PN}.}   
\label{fig:fig2} 
\end{figure} 

\begin{figure} 
\includegraphics[width=1.0\linewidth]{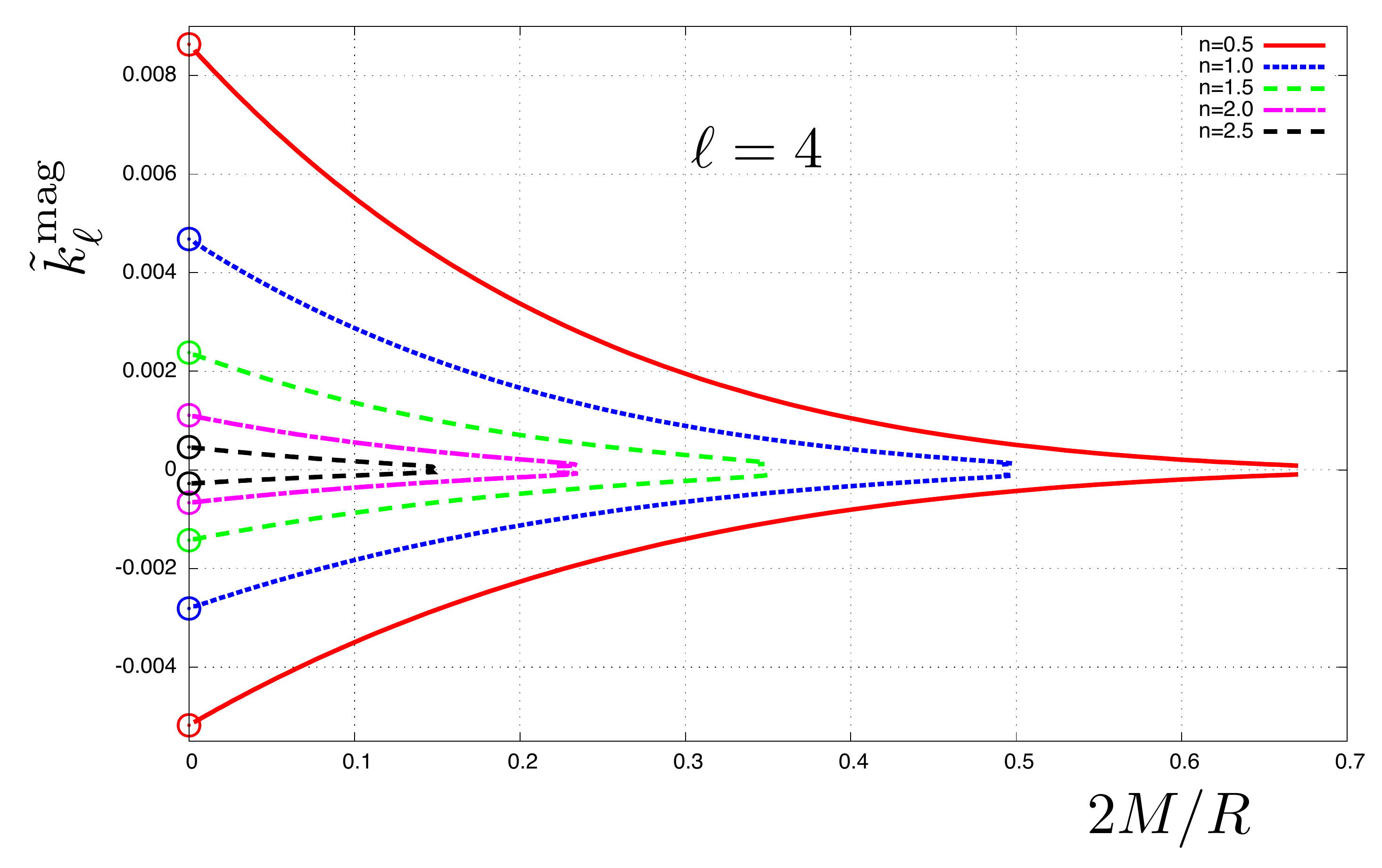}
\caption{Rescaled gravitomagnetic Love number 
$\tilde{k}^{\rm mag}_\ell$ for $\ell = 4$. The caption of
Fig.~\ref{fig:fig1} provides additional details.}  
\label{fig:fig3} 
\end{figure} 

\begin{figure} 
\includegraphics[width=1.0\linewidth]{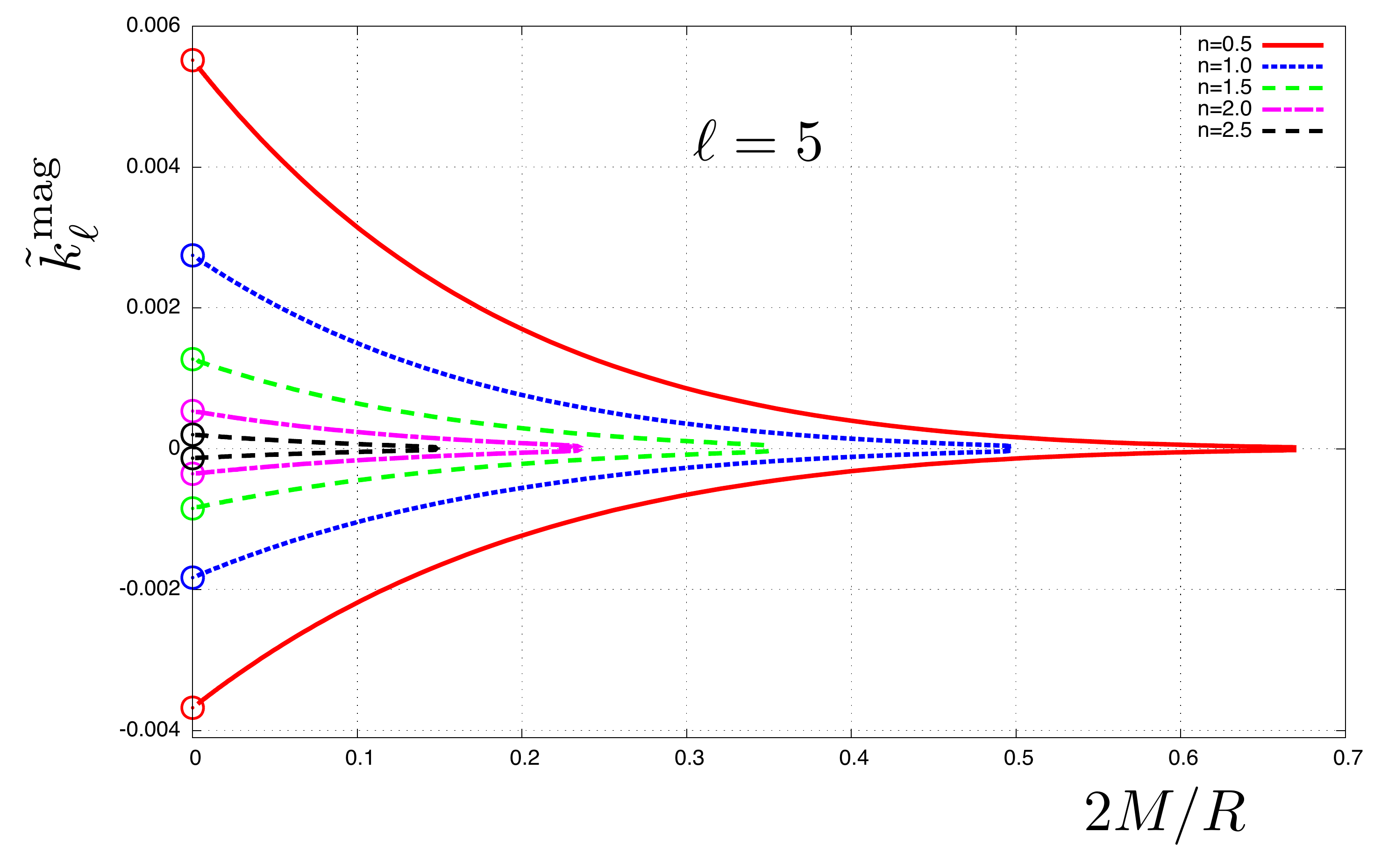}
\caption{Rescaled gravitomagnetic Love number 
$\tilde{k}^{\rm mag}_\ell$ for $\ell = 5$. The caption of
Fig.~\ref{fig:fig1} provides additional details.}  
\label{fig:fig4} 
\end{figure} 

Equation (\ref{deq1}) is to be integrated outward from $r = 0$, near
which the solution behaves as $h_t^{\ell m} \propto r^{\ell+1}$. The
internal solution is matched at $r=R$ to the external solution
provided by Binnington and Poisson \cite{binnington-poisson:09},  
\begin{equation} 
h_t^{\ell m} = \frac{2}{3(\ell-1)\ell} r^{\ell+1}     
\biggl[ A_4 - 2 \frac{\ell+1}{\ell} K^{\rm mag}_\ell (2M/r)^{2\ell+1} 
B_4 \biggr] \B^{\ell m}, 
\label{ht_ext} 
\end{equation}  
where   
\begin{subequations} 
\begin{align} 
A_4 &:= F(-\ell+1,-\ell-2;-2\ell;2M/r), \\ 
B_4 &:= F(\ell-1,\ell+2;2\ell+2;2M/r) 
\end{align} 
\end{subequations} 
with $F(a,b;c;x)$ denoting the hypergeometric function, 
and $\B^{\ell m}$ is the spherical-harmonic packaging of the
tidal moments $\B_{j_1 j_2 \cdots j_\ell}$ introduced in
Eq.~(\ref{g0j}), defined by 
\begin{equation} 
\B_{j_1 j_2 \cdots j_\ell} \Omega^{j_1} \Omega^{j_2} \cdots
\Omega^{j_\ell} =  \sum_{\ell = -m}^\ell \B^{\ell m} Y^{\ell m}, 
\label{STF_Ylm} 
\end{equation} 
in which $\Omega^j := [\sin\theta\cos\phi, \sin\theta\sin\phi,
\cos\theta]$. 

A practical formulation of this procedure was provided by Landry and
Poisson \cite{landry-poisson:14}. It involves introducing the
logarithmic derivative $ \kappa_\ell := d \ln h_t^{\ell m}/d\ln r$,
which satisfies the nonlinear differential equation  
\begin{equation} 
r \frac{d\kappa_\ell}{dr} 
+ \kappa_\ell (\kappa_\ell - 1) - P \kappa_\ell - Q = 0.  
\label{deq2} 
\end{equation} 
This equation is integrated from $r=0$, at which 
$\kappa_\ell = \ell + 1$, to $r = R$, at which 
$\kappa_\ell = \kappa_\ell^{\rm s}$. The matching condition at $r=R$
produces    
\begin{equation} 
K_\ell^{\rm mag} = \biggl( \frac{R}{2M} \biggr)^{2\ell} 
\tilde{k}^{\rm mag}_\ell 
\end{equation} 
with 
\begin{equation} 
\tilde{k}^{\rm mag}_\ell  =  \frac{\ell}{2(\ell+1)} \frac{R}{2M} 
\frac{ RA'_4 - (\kappa_\ell^{\rm s} - \ell - 1) A_4 }
{ RB'_4 - (\kappa_\ell^{\rm s} + \ell) B_4 },  
\label{ktilde} 
\end{equation} 
in which a prime indicates differentiation with respect to $r$. The
overall scaling of $K^{\rm mag}_\ell$ with $(R/2M)^{2\ell}$ differs
from the convention adopted in Damour and Nagar \cite{damour-nagar:09}
and Binnington and Poisson \cite{binnington-poisson:09}, who
introduced instead a scalefree Love number $k_\ell^{\rm mag}$ via the
relation $K_\ell^{\rm mag} = (R/2M)^{2\ell+1} k_\ell^{\rm mag}$. The
missing factor of $R/2M$ was incorporated in Eq.~(\ref{ktilde}), and
it ensures that $\tilde{k}_\ell^{\rm mag}$ tends to a nonzero value in
the limit $2M/R \to 0$; this is unlike $k_\ell^{\rm mag}$, which goes
to zero in the limit.  

In Figs.~\ref{fig:fig1}--\ref{fig:fig4} we present sample calculations
of the rescaled gravitomagnetic Love numbers 
$\tilde{k}_\ell^{\rm mag}$ for a polytropic equation of state  
$p = K \rho^{1+1/n}$, $\epsilon = n p$, where $K$ and $n$ are
constants. The details of integrating Eq.~(\ref{deq2})
for this specific case are presented in Appendix \ref{sec:polytropes}. We
see from the figures that the tidal response of an irrotational body
is dramatically different from the response of a body in strict
hydrostatic equilibrium: the gravitomagnetic Love numbers of an
irrotational body are negative, while they are positive for a body in
strict hydrostatic equilibrium. The change of sign can be seen to
originate from the $8\pi r^2 (\mu+p)$ term in $Q$. Both sets of Love 
numbers, however, share  the properties that they decrease (in
absolute value) with increasing $2M/R$ and with increasing polytropic
index $n$; because increasing $n$ decreases $p$ for a given $\rho$,
and therefore produces a body that is more centrally dense, both
properties are associated with the fact that a more compact body
develops smaller multipole moments.    

\section{Post-Newtonian calculation of the gravitomagnetic Love
  numbers} 
\label{sec:PN} 

In this section we exploit the methods of post-Newtonian theory to
calculate the gravitomagnetic response of a weakly self-gravitating
body to an applied tidal field. Our calculation returns an expression  
for $K^{\rm mag}_\ell$ that reproduces the leading-order behavior of 
the relativistic function when the body's compactness 
$2GM/(c^2 R)$ is small. In this section we restore powers of $G$ and
$c$, which were set equal to unity in the preceding sections.   

The post-Newtonian metric is expressed in harmonic coordinates 
$(ct, x^j)$ as 
\begin{equation} 
g_{00} = -1 + \frac{2}{c^2} U + O(c^{-4}), \qquad 
g_{0j} = -\frac{4}{c^3} U_j + O(c^{-7}), \qquad 
g_{jk} = \delta_{jk} \biggl( 1 + \frac{2}{c^2} U \biggr) 
+ O(c^{-4}),  
\label{PNmetric} 
\end{equation} 
and it involves a vector potential $U_j$ in addition to the familiar
Newtonian potential $U$, which satisfies the Poisson equation
$\nabla^2 U = -4\pi G \rho$. The post-Newtonian terms of order
$c^{-4}$ in $g_{00}$ are not required in our developments, and the
second post-Newtonian terms of order $c^{-5}$ in $g_{0j}$ are absorbed
into $U_j$. 

As was described in Sec.~\ref{sec:intro}, the tidal environment is
taken to be a pure gravitomagnetic $\ell$-pole field characterized by
the symmetric-tracefree (STF) tensor $\B_{k_1 k_2 \cdots k_\ell}(t)$,
which is assumed to vary slowly with time, so that we may neglect time
derivatives in the field equations. The tidal field is described by
the vector potential   
\begin{equation} 
U_j^{\rm tidal} = -\frac{1}{6(\ell-1)} r^\ell 
\epsilon_{jpq} \Omega^p \B^q_{\ k_2 k_3 \cdots k_\ell} 
\Omega^{k_2} \Omega^{k_3} \cdots \Omega^{k_\ell}, 
\label{Uj_tidal} 
\end{equation}  
where $\epsilon_{jpq}$ is the completely antisymmetric permutation
symbol, $r^2 := \delta_{jk} x^j x^k$, and $\Omega^j := x^j/r$.  
The STF nature of the tidal moment $\B_{k_1 k_2 \cdots k_\ell}(t)$
ensures that the vector potential satisfies Laplace's equation
$\nabla^2 U_j^{\rm tidal} = 0$ as well as the harmonic gauge condition
$\partial^j U_j^{\rm tidal} = 0$. It also implies that the vector
potential can be expressed in the alternative form 
\begin{equation} 
U_j^{\rm tidal} = -\frac{1}{6(\ell-1)} r^\ell
\epsilon_{jpq} \B^q_{\ k_2 k_3 \cdots k_\ell} 
\Omega^\stf{p k_2 k_3 \cdots k_\ell},  
\end{equation}   
in which $\Omega^{k_1 k_2 \cdots k_\ell} := \Omega^{k_1} \Omega^{k_2} 
\cdots \Omega^{k_\ell}$ and the angular brackets indicate the
operation of trace-removal. It is easy to show that a transformation
to spherical polar coordinates $(r,\theta^A)$ brings the metric to the
form displayed in Eq.~(\ref{ht_ext}), in which we set $A_4 = 1 +
O(c^{-2})$ to respect the post-Newtonian approximation, and $B_4 = 0$
because the body's response has not yet been incorporated within the
vector potential.   

The body's response to the applied tidal field is governed by the
post-Newtonian Euler equation and the field equation satisfied by the
vector potential. The relevant aspects can be collected from the
textbook by Poisson and Will \cite{poisson-will:14}. From their
Eq.~(7.15b), in which we neglect the time derivatives, and their
Eq.~(7.23b) we obtain   
\begin{equation} 
U_j = U_j^{\rm tidal} + G \int_\M 
  \frac{s_j(\bm{x'})}{|\bm{x}-\bm{x'}|}\, d^3x' 
\label{Uj_resp} 
\end{equation} 
in which the domain of integration $\M$ is truncated to $r' < \R$,
where $\R$ is a cutoff radius to be specified below. The body's
response is captured by the Poisson integral, and the integrand
involves the effective current density $s^j := c^{-1} \tau^{0j}$
which incorporates both a matter and field contribution. An expression
for this is displayed in Exercise 8.4 of Poisson and Will; we have    
\begin{equation} 
s^j = \rho v^j \bigl[ 1 + O(c^{-2}) \bigr]   
+ \frac{1}{\pi G c^2} \bigl( \partial^j U^k - \partial^k U^j
\bigr) \partial_k U + O(c^{-4}), 
\label{current} 
\end{equation} 
in which we neglect a term involving $\partial_t U$ in the field  
contribution to the mass current, as well as the $O(c^{-2})$ terms 
multiplying $\rho v^j$ because, as we shall see presently, the
velocity field is itself of order $c^{-2}$. Our expression for $U_j$
should be multiplied by the factor $(1 - 2U/c^2)$ arising from
Eq.~(7.23b) of Poisson and Will. This factor can be neglected when it
multiplies the Poisson integral, but it produces a relativistic
correction to $U_j^{\rm tidal}$. We shall not be interested in such
corrections, and choose to drop the factor all together.      

The velocity field is determined by the irrotational condition that we
wish to place on the fluid. Writing $u^\alpha = \gamma(c, v^j)$ for
the four-velocity, in which $\gamma$ is a normalization factor, a
short computation reveals that the circulation of a post-Newtonian
fluid around any spatial circuit $c$ is given by 
\begin{equation} 
C(c) = \oint_c \gamma h \biggl[ 
\biggl( 1 + \frac{2}{c^2} U \biggr) v_j 
- \frac{4}{c^2} U_j + O(c^{-4}) \biggr] dx^j. 
\end{equation} 
An irrotational state therefore requires 
\begin{equation} 
v_j = \frac{4}{c^2} U_j + O(c^{-4}). 
\label{vPN} 
\end{equation} 
This expression agrees with results previously obtained by Shapiro
\cite{shapiro:96} and Favata \cite{favata:06}.  

The calculation of the body's response proceeds by inserting
Eq.~(\ref{vPN}) within Eq.~(\ref{current}), and replacing $U_j$ by
$U_j^{\rm tidal} + O(c^{-2})$ in this expression. We find that the
velocity contribution to the mass current is given by 
\begin{equation} 
s^1_j = -\frac{2\lambda}{3(\ell-1) c^2} \rho r^\ell 
\epsilon_{jpq} \B^q_{\ k_2 k_3 \cdots k_\ell} 
\Omega^\stf{p k_2 k_3 \cdots k_\ell}, 
\label{s1} 
\end{equation} 
and that the field contribution is 
\begin{equation} 
s^2_j = -\frac{\ell+1}{6 \pi (\ell-1) c^2} m r^{\ell-3}  
\epsilon_{jpq} \B^q_{\ k_2 k_3 \cdots k_\ell} 
\Omega^\stf{p k_2 k_3 \cdots k_\ell}. 
\label{s2} 
\end{equation} 
To arrive at the last expression we recalled the fact that for a
spherical body, $\partial_k U = -(Gm/r^2) \Omega_k$, in which $m(r)$  
is the mass inside a sphere of radius $r$. We also inserted a factor
of $\lambda \equiv 1$ in Eq.~(\ref{s1}) to again track the influence
of the internal motions on the body's gravitomagnetic response.    

The next step is to evaluate the Poisson integral in
Eq.~(\ref{Uj_resp}).  The calculation requires the addition theorem
for spherical harmonics,  
\begin{equation} 
\frac{1}{|\bm{x}-\bm{x'}|} = \sum_{\ell m} \frac{4\pi}{2\ell + 1} 
\frac{r_<^\ell}{r_>^{\ell+1}} Y^*_{\ell m}(\theta',\phi')
Y_{\ell m}(\theta,\phi), 
\label{addition} 
\end{equation} 
in which $r_< = \mbox{min}(r,r')$ and $r_> = \mbox{max}(r,r')$,  
and the identity displayed in Eq.~(1.171) of Poisson and Will, 
\begin{equation} 
\sum_m Y_{\ell m}(\theta,\phi) \int Y^*_{\ell m}(\theta',\phi') 
\Omega^{\prime \stf{k_1 k_2 \cdots k_{\ell'}}}\, d\Omega' 
= \delta_{\ell\ell'} 
\Omega^{\stf{k_1 k_2 \cdots k_{\ell}}}.   
\label{identity} 
\end{equation} 
Evaluation of the Poisson integral for $s^1_j$ is straightforward,
and when $r > R$ we obtain 
\begin{equation} 
U^1_j = -\frac{1}{6(\ell-1)} r^\ell
\epsilon_{jpq} \Omega^p \B^q_{\ k_2 k_3 \cdots k_\ell} 
\Omega^{k_2} \Omega^{k_3} \cdots \Omega^{k_\ell} 
\biggl[ \frac{16\lambda}{2\ell+1} \beta_\ell \frac{GM}{c^2}
\frac{R^{2\ell}}{r^{2\ell+1}} \biggr], 
\end{equation} 
in which 
\begin{equation} 
\beta_\ell := \frac{\pi}{M R^{2\ell}} \int_0^R \rho r^{2\ell+2}\, dr 
\label{beta} 
\end{equation} 
is a dimensionless moment of the mass density. In this case the domain
of integration is naturally limited to the volume occupied by the
body, and the cutoff radius $\R$ is irrelevant. It does, however, play
a role in the Poisson integral for $s^2_j$, which features the radial 
integral  
\begin{equation} 
J := \int_0^{\R} m(r') r^{\prime \ell - 1}
\frac{r_<^\ell}{r_>^{\ell+1}}\, dr'. 
\end{equation} 
To evaluate this for $r > R$ we break up the integration domain into a 
first segment $0 < r' < R$ in which $r_< = r'$, $r_> = r$, and $m(r')$
is unspecified, a second segment $R < r' < r$ in which $r_< = r'$,
$r_> = r$, and $m(r')=M$, and a third segment $r < r' < \R$ in which
$r_< = r$, $r_> = r'$, and $m(r')=M$. We obtain 
\begin{equation} 
J = \frac{1}{r^{\ell+1}} \biggl[ \int_0^R m(r) r^{2\ell-1}\, dr 
- \frac{1}{2\ell} M R^{2\ell} \biggr] 
+ \frac{2\ell+1}{2\ell} M r^{\ell - 1}  
- \frac{M}{\R} r^\ell. 
\end{equation} 
To simplify this we integrate the first term by parts, making use of
the Newtonian field equation $dm/dr = 4\pi r^2 \rho$. We also discard
the term proportional to $M r^{\ell - 1}$, because it merely gives
rise to an uninteresting correction of order $GM/(c^2 r)$ to the tidal
potential of Eq.~(\ref{Uj_tidal}). The term proportional to 
$M r^\ell/\R$ can be seen to alter the amplitude of the tidal
potential by a correction of order $GM/(c^2 \R)$, and we eliminate
this meaningless shift by setting $\R = \infty$. With all this we find
that  
\begin{equation} 
J = -\frac{2}{\ell} \beta_\ell \frac{M R^{2\ell}}{r^{\ell + 1}}, 
\end{equation} 
and completing the evaluation of the Poisson integral, we arrive at  
\begin{equation}     
U^2_j = -\frac{1}{6(\ell-1)} r^\ell
\epsilon_{jpq} \Omega^p \B^q_{\ k_2 k_3 \cdots k_\ell} 
\Omega^{k_2} \Omega^{k_3} \cdots \Omega^{k_\ell} 
\biggl[ -\frac{8(\ell+1)}{\ell(2\ell+1)} \beta_\ell \frac{GM}{c^2}
\frac{R^{2\ell}}{r^{2\ell+1}} \biggr].
\end{equation} 
Adding the contributions, we find that the body's gravitomagnetic
response to the applied tidal field is described by 
\begin{equation} 
U^{\rm resp}_j = -\frac{1}{6(\ell-1)} r^\ell
\epsilon_{jpq} \Omega^p \B^q_{\ k_2 k_3 \cdots k_\ell} 
\Omega^{k_2} \Omega^{k_3} \cdots \Omega^{k_\ell} 
\biggl[  \frac{8[(2\lambda-1)\ell - 1]}{\ell(2\ell+1)} \beta_\ell
\frac{GM}{c^2} \frac{R^{2\ell}}{r^{2\ell+1}} \biggr]. 
\end{equation}  

The complete vector potential is $U^{\rm tidal}_j 
+ U^{\rm resp}_j$, and comparison with Eq.~(\ref{g0j}) reveals 
the post-Newtonian expression for the gravitomagnetic Love
numbers. Recalling Eq.~(\ref{beta}), we find that 
\begin{equation} 
K^{\rm mag}_\ell = \biggl( \frac{R}{2GM/c^2} \biggr)^{2\ell}
\tilde{k}^{\rm mag}_\ell 
\end{equation} 
with 
\begin{equation} 
\tilde{k}^{\rm mag}_\ell := 
-\frac{2 \bigl[(2\lambda-1)\ell - 1 \bigr]}{(\ell+1)(2\ell+1)} 
\frac{\pi}{M R^{2\ell}} \int_0^R \rho r^{2\ell+2}\, dr. 
\label{ktildePN} 
\end{equation} 
We recall that $\lambda$ keeps track of the internal motions
associated with the fluid's irrotational state. Setting $\lambda = 1$
places the body in the irrotational state, and we observe that 
{\it negative} gravitomagnetic Love numbers must be 
assigned to such a body, irrespective of the multipole order 
$\ell \geq 2$ and the equation of state. By contrast, setting $\lambda
= 0$ places the body in a strict hydrostatic equilibrium, and such a
body necessarily comes with {\it positive} gravitomagnetic Love
numbers. These properties were featured in the fully relativistic
results displayed in Figs.~\ref{fig:fig1}--\ref{fig:fig4}, and indeed,
we observe that the post-Newtonian expression of Eq.~(\ref{ktildePN})
accurately reproduces the relativistic Love numbers in the limit
$2GM/(c^2 R) \to 0$.   

Our post-Newtonian calculation of the gravitomagnetic Love numbers
completes previous attempts carried out by Favata \cite{favata:06} and
Damour and Nagar \cite{damour-nagar:09}. In his work (see Sec.~III B
of his paper), Favata introduces a definition for the Love numbers
that accounts only for the velocity term $s^1_j$ in the effective mass
current; it omits the field term $s^2_j$, and Favata therefore
produces only the $\lambda$ term in Eq.~(\ref{ktildePN}). On the other
hand, the post-Newtonian calculation of Damour and Nagar 
(see Sec.~VIII of their paper) places the fluid in a strict
hydrostatic equilibrium instead of the irrotational state, and
therefore accounts only for the $\lambda$-independent term in
Eq.~(\ref{ktildePN}). (It should be noted that the Damour-Nagar
definition for the gravitomagnetic Love numbers includes a
multiplicative minus sign compared to ours; their Love numbers are
negative when ours are positive.) Our own calculation brings the two
partial stories together, and generalizes the previous calculations
(which were limited to $\ell=2$) to arbitrary multipole order $\ell$.    

Equation (\ref{ktildePN}) can be evaluated in closed form in a few
simple cases. For a constant density body we find that 
\begin{equation} 
\tilde{k}^{\rm mag}_\ell = 
-\frac{3\bigl[ (2\lambda-1)\ell - 1 \bigr]}{2(\ell+1)(2\ell+1)(2\ell+3)}. 
\end{equation} 
For an $n=1$ Newtonian polytrope, for which $\rho = M/(4R^2 r)
\sin(\pi r/R)$, we have that 
\begin{subequations} 
\begin{align} 
\tilde{k}^{\rm mag}_2 &= 
-\frac{(4\lambda-3)(\pi^4-20\pi^2+120)}{20 \pi^4}
\simeq (-0.0068498, 0.020549), \\ 
\tilde{k}^{\rm mag}_3 &= 
-\frac{(3\lambda-2)(\pi^6-42\pi^4+840\pi^2-5040)}{28 \pi^6}
\simeq (-0.0044829, 0.0089658), \\ 
\tilde{k}^{\rm mag}_4 &= 
-\frac{(8\lambda-5)(\pi^8-72\pi^6+3024\pi^4-60480\pi^2+362880)} 
  {90\pi^8}
\simeq (-0.0028102, 0.0046836), \\ 
\tilde{k}^{\rm mag}_5 &= 
-\frac{(5\lambda-3)(\pi^{10}-110\pi^8+7920\pi^6-332640\pi^4
  +6652800\pi^2-39916800)}{66\pi^{10}}
\simeq (-0.0018302, 0.0027454). 
\end{align} 
\end{subequations}   
The numerical values correspond to $\lambda = 1$ and $\lambda = 0$,
respectively. 

\begin{acknowledgments} 
One of us (EP) is grateful to the Canadian Institute of Theoretical
Astrophysics for its warm hospitality during a research leave from the 
University of Guelph. This work was supported by the Natural Sciences
and Engineering Research Council of Canada.     
\end{acknowledgments}    

\appendix
\section{Structure and perturbation equations for polytropes}     
\label{sec:polytropes} 

To perform the computations described in Sec.~\ref{sec:Love} for the  
polytropic equation of state $p = K \rho^{1+1/n}$, $\epsilon = n p$ we
recast the background field equations of Sec.~\ref{sec:unperturbed}
and Eq.~(\ref{deq2}) in convenient, dimensionless forms. To achieve
this we introduce the central density $\rho_c := \rho(r=0)$, the
central pressure $p_c := p(r=0)$, the length scale $r_0$ defined by  
$r_0^2 := (n+1) p_c/(4\pi\rho_c^2)$, and the mass scale 
$m_0 := 4\pi \rho_c r_0^3$. A useful dimensionless parameter is 
$b := p_c/\rho_c = K \rho_c^{1/n}$, which can act as a substitute for
the central density as a label of polytropic models. A frequently
encountered combination of scaling quantities is $m_0/r_0 = (n+1) b$.  

We next introduce the dimensionless, Lane-Emden-type variables $\xi$,
$\theta$, and $\nu$, such that $r = r_0 \xi$, $\rho = \rho_c \theta^{n}$,  
$p = p_c \theta^{n+1}$, $\mu = \rho_c \theta^n (1+nb\theta)$, and 
$m = m_0 \xi^3 \nu$. The equations that determine the internal
structure of the unperturbed polytrope are then  
\begin{equation} 
\xi \frac{d\nu}{d\xi} = \theta^n(1 + n b \theta) - 3\nu 
\end{equation} 
and 
\begin{equation} 
\xi \frac{d\theta}{d\xi} = -\xi^2 f^{-1}  
\bigl[ 1 + (n+1) b \theta \bigr] \bigl( \nu + b\theta^{n+1} \bigr), 
\end{equation} 
with $f = 1 - 2(n+1) b \xi^2 \nu$. An equation can also be
displayed for the gravitational potential $\psi$, but this is not
needed to calculate the Love numbers. The integrations begin at 
$\xi = 0$ with $\theta(0)  = 1$ and $\nu(0) = \frac{1}{3}(1+nb)$. They
proceed until $\xi = \xi_{\rm s}$ at which $\theta = 0$ and 
$\nu = \nu_{\rm s}$. The body's compactness can then be calculated as  
$2M/R = 2(n+1) b \xi^2_{\rm s} \nu_{\rm s}$. In the limit 
$b \to 0$ the equations reduce to the standard Lane-Emden form.  

The dimensionless version of Eq.~(\ref{deq2}) is
\begin{equation} 
\xi \frac{d\kappa_\ell}{d\xi} + \kappa_\ell (\kappa_\ell-1) 
- P\kappa_\ell - Q = 0, 
\end{equation} 
with 
\begin{subequations} 
\begin{align} 
P &= (n+1) b \xi^2 \theta^n f^{-1} 
\bigl[ 1 + (n+1) b \theta \bigr], \\ 
Q &= \frac{1}{f} \biggl\{ \ell(\ell+1) - 2(n+1) b \xi^2 
\Bigl[ 2\nu + (2\lambda-1) \theta^n \bigl( 1 + (n+1) b \theta \bigr)
\Bigr] \biggr\}. 
\end{align} 
\end{subequations} 
The integration begins at $\xi = 0$ with $\kappa_\ell = \ell+1$ and 
proceeds until $\xi = \xi_{\rm s}$ at which 
$\kappa_\ell = \kappa_\ell^{\rm s}$. 

In practice it is helpful to use $x := \ln \xi$ as the independent
variable, and to start the integration at a large, negative value of
$x$. Starting values for $\theta$, $\nu$ and $\kappa_\ell$ can be
obtained from the Taylor expansions $\theta = 1 + \theta_2 \xi^2 
+ \theta_4 \xi^4 + \cdots$, $\nu = \frac{1}{3}(1+nb) + \nu_2 \xi^2 
+ \nu_4 \xi^4 + \cdots$, and $\kappa_\ell = \ell+1 
+ \kappa_{\ell,2} \xi^2 + \kappa_{\ell,4} \xi^4 + \cdots$, in which
the various coefficients can be determined from the differential
equations.   

\bibliography{../bib/master} 
\end{document}